\documentstyle[11pt,aasms4]{article}   

\slugcomment{ApJ Letters, accepted for publication, 1997 August 12th}

\lefthead{Belloni T. et al.}

\begin{document}

\def\spose#1{\hbox to 0pt{#1\hss}}
\def\lta{\mathrel{\spose{\lower 3pt\hbox{$\mathchar"218$}}
     \raise 2.0pt\hbox{$\mathchar"13C$}}}
\def\gta{\mathrel{\spose{\lower 3pt\hbox{$\mathchar"218$}}
     \raise 2.0pt\hbox{$\mathchar"13E$}}}
\def\Msun{{\rm M}_\odot}
\def\msun{{\rm M}_\odot}
\def\Rsun{{\rm R}_\odot}
\def\Lsun{{\rm L}_\odot}
\def\half{{1\over2}}
\def\RL{R_{\rm L}}
\def\zs{\zeta_{s}}
\def\zR{\zeta_{\rm R}}
\def\dJJ{{\dot J\over J}}
\def\dMM{{\dot M_2\over M_2}}
\def\tKH{t_{\rm KH}}
\def\eck#1{\left\lbrack #1 \right\rbrack}
\def\rund#1{\left( #1 \right)}
\def\wave#1{\left\lbrace #1 \right\rbrace}
\def\dd{{\rm d}}


\title{A unified model for the spectral variability in GRS~1915+105}
\author{
        T.~Belloni\altaffilmark{1},
        M.~M\'endez\altaffilmark{1,2},
        A.R.~King\altaffilmark{3},
        M.~van der Klis\altaffilmark{1},
        J.~van Paradijs\altaffilmark{1,4}
       }

\altaffiltext{1} {Astronomical Institute ``Anton Pannekoek'',
       University of Amsterdam and Center for High-Energy Astrophysics,
       Kruislaan 403, NL-1098 SJ Amsterdam, the Netherlands}

\altaffiltext{2}{Facultad de Ciencias Astron\'omicas y Geof\'{\i}sicas, 
       Universidad Nacional de La Plata, Paseo del Bosque S/N, 
       1900 La Plata, Argentina}

\altaffiltext{3}{Astronomy Group, University of Leicester,
                 Leicester LE1 7RH, United Kingdom}

\altaffiltext{4}{Physics Department, University of Alabama in Huntsville,
       Huntsville, AL 35899, USA}

\begin{abstract}
We have analyzed the spectral variations of the superluminal black-hole
X-ray binary GRS~1915+105 by using data obtained with the PCA on the
Rossi XTE. We find that, despite the marked differences in the structure
and the time scale of variability, all spectral changes can be attributed
to the rapid disappearing of the inner region of an
accretion disk, followed by a slower re-filling of the emptied
region. The time scale for each event is determined by the extent of the
missing part of the disk. The observed relation between the duration of
an event and the radius of the disappearing region matches remarkably
well the expected radius dependence of the viscous time scale for the
radiation-pressure dominated region of an accretion disk.
\end{abstract}

\keywords{accretion, accretion disks ---
          binaries: close --- black hole physics -- instabilities
          -- X-rays: stars --- stars: individual GRS~1915+105}

\section{INTRODUCTION}

The X--ray source GRS~1915+105 is the best of the two only examples
of galactic objects that show
superluminal expansion in radio observations (\cite{mr94}).
It is located at a distance of 12.5 kpc and its relativistic jets are
directed at 70$^\circ$ from the line of sight (\cite{mr94}).
After its discovery as an X-ray transient in 1992 (\cite{cas92}), a number
of outbursts has been reported, although it is possible that the source
has never completely switched off in between.
The Rossi X-ray Timing Explorer (RXTE) started observing GRS~1915+105
in a bright and variable state in April 1996 and continued to observe it
regularly at least once per week since then. During this period, the source
has displayed a remarkable richness in variability, ranging from
quasi-periodic burst-like events, deep regular dips and strong
quasi-periodic oscillations, alternated with quiescent periods 
(\cite{gmr96},\cite{mrg96},\cite{cst97},\cite{bmk97},\cite{tcs97}).
Because of its similarities with the other galactic superluminal source
GRO~1655-40 (\cite{zh94}), whose dynamical mass estimate implies the presence 
of a black hole (\cite{bai95}), and because of its high luminosity exceeding
the Eddington limit for a neutron star, the source is suspected
to host a black hole.

In a previous paper
(\cite{bmk97}, hereafter Paper I) 
we showed that the complicated light curve
of GRS~1915+105 can be described by the rapid ($\sim$1 s) appearance and
disappearance of emission from an optically thick inner accretion disk.
In this paper we report on the results of the analysis of a 
later observation which allowed a more detailed investigation of the
spectral variability of the source. We find that the variation 
time scale $t_{\rm rec}$ of the disk source is completely specified by the
maximum size $R_{\rm max}$ of it inner radius. The relation between these 
quantities is precisely as expected for the viscous evolution of a 
radiation--pressure dominated accretion disk.


\section{X--RAY OBSERVATIONS}

Among the many PCA observations of GRS~1915+105, that of
1997 June 18th deserves particular
attention, since it reproduces within one day most of the timing behaviors
observed from the source.  The data consist of two segments of roughly one hour each.
Since the two segments are essentially similar, we will restrict ourselves 
to the second one, which starts at 14:36 UT and ends at 15:35 UT. 
A 1200 s representative stretch of the 2.0-40.0 keV light curve, 
binned to 1 second, can 
be seen in the top panel of Figure 1. It consists of a sequence of `bursts'
of different duration with quiescent intervals in between. All bursts start
with a well-defined sharp peak and decay faster than they rise. The longer
bursts show oscillation (or sub-bursts) towards the end.
In the following we define an event as a full cycle of one quiescent
interval followed by one burst. With this
definition there is no interval between events. In order to maintain 
precision, separate events are considered only as those whose quiescent 
count rate goes below 10000 cts/s in the total 2.0--40.0 keV band.  
We measured the start
of an event (corresponding to the end of the previous one) as the time
of the small dip at the end of the decay (see Figure 1). Since all the
bursts start with a sharp peak, the time of the peak can be taken as
the separation between the quiescent phase and the burst. Finally, the bottom
level of each quiescent part has been measured as the lowest level excluding the
small dip mentioned above. With these numbers it is possible to associate
four quantities to an event: total length, length of the quiescent part, length
of the outburst and level of the bottom. Two strong correlations emerge:
(i) the longer the event, the lower the bottom level (see below); and (ii) a 
one-to-one relation between quiescent and burst duration (Figure 2). 
No correlation is seen between a burst duration and the following quiescent 
interval.

In order to follow the spectral evolution, we produced two X-ray colors:
HR1 = B/A and HR2 = C/(A+B), where A, B and C are the counts in the 
2.0--5.0 keV, 5.0--13.0 keV and 13.0--40.0 keV intervals respectively.
We produced color-color diagrams by selecting events according to their
total length. It turns out that there is a very strong correlation
between the pattern traced out in the CC diagram by each event, and its
duration.
In Figure 3 one can see two extreme cases and an intermediate
one. All other events follow a similar pattern. During short events
(top panel) the quiescent period is at the bottom left in the C-C diagram, 
then the source moves up to the top left during peak, then down and right 
to the dip. In longer events the sequence
is the same, with the only difference that the quiescent part is located 
farther to the right the longer the event is. 
The small oscillations at the end of long bursts show up as
oscillations around the peak points in the C-C diagram (mid and bottom
panels in Figure 3). 

The spectral evolution is extremely reproducible throughout the
different events. We therefore divided the C-C diagram in regions 0.01 wide
in HR2 and 0.1 wide in HR1 and accumulated energy spectra for all the
(74) populated regions of the diagram.
The spectra were accumulated in 48 energy bins with 1 second time 
resolution. The background subtraction was performed with a background
spectrum extracted from a blank sky observation and normalized to the
highest channels where the source signal is absent.
We used the latest detector
response matrix available (April 1997, Version 2.1.2). A systematic error
of 2\% has been added to the data (see Paper I). In order to avoid
possible problems due to the background subtraction, we limited our fits
to energies below 30.0 keV.
Following paper I, for each of the regions we 
fitted the data with a ``standard'' spectral model
for black-hole candidates, consisting of a disk-blackbody model 
(\cite{mal84}) plus
a power law. A broad Gaussian line was added at a fixed energy of 6.3 keV
as in Paper I. We found the interstellar absorption to be
stable around 4.5$\times 10^{22}$ cm$^{-2}$, so it was fixed to that 
value. Notice that this is lower than the value found in paper I because of 
the improved calibration. It is slightly lower than the best-fit ROSAT 
value (\cite{gmr96}). The resulting parameters for the power law component
depend on the inclusion of the Gaussian line mentioned above,
but the disk-blackbody parameters are independent of this. 
In the following we will restrict ourselves to the discussion of the latter.
Since both distance and inclination of the system are known (the latter
is assumed to be equal to the observed inclination of the radio jet, see
\cite{mr94}), it
is possible to derive the inner radius of the accretion disk directly
from the model without further assumptions.

The fitting procedure yielded spectral parameters for each of the 74
regions. By identifying the region of each 1-s bin, it is possible to
produce a time history of relevant spectral parameters with a resolution
of 1 second. The inner radius and temperature are shown in the bottom
two panels of Figure 1. As it can be seen, during bursts the temperature is 
above 2 keV and the radius stable around 20 km, parameters consistent
with those reported in Paper I. During quiescent phases, the temperature
drops to lower values and the radius increases.  There is a strong correlation
between the length of the event and the largest radius reached (Figure 4).
In order to check the results of our `grid' procedure, we extracted
spectra for the longest event in 16 s bins from the Standard~2 data, 
which have a much better energy resolution. We obtained similar results.
Since the spectral parameters coming from these data are more accurate,
we used those for the estimate of the values of the accretion rate.
Notice that the radii determined for the disk hole are smaller than that
reported in Paper I, primarily due to the new PCA calibrations.
Because of these differences, the behavior of the accretion rate derived
from the model fits (which depends on the third power of the inner radius)
is now radically different: the value of $\dot M$ during the quiescent phase is
around 1.7--10$\times 10^{-8}$M$_\odot$/yr (the range being determined 
by the spin of the central black hole, which 
determines the innermost stable orbit), while
in the burst phase it is around a factor of two higher and variable.

We produced similar color-color diagrams for a number of other RXTE observations
of GRS~1915+105. All observations we analyzed follow the pattern outlined
above, except that of 1996 June 16th (\cite{gmr96}).
All the quasi-periodic bursts observed in many
observations (see \cite{tcs97}) are consistent with
repetitive short events like the ones described here.

\section{DISCUSSION}

The results described in the previous section can be interpreted within 
the model presented in Paper I, providing a unified picture
of the variability observed in GRS~1915. In Paper I, we modeled the 
large amplitude changes as emptying and replenishing of the inner 
accretion disk caused by a viscous-thermal instability. The small radius
observed during the quiescent period was identified with the innermost stable
orbit around the black hole, while the large radius during the burst phase
was the radius of the emptied section of the disk. The smaller
oscillations were interpreted as failed attempts to empty the inner disk.
As it can be seen from Figure 1, from this observation we find 
that all variations, from major events
like the ones described in Paper I to small oscillations observed at the
end of a large event, can be modeled in exactly the same fashion (see
Figure 1). 
In this scenario, the ``flare state'' presented in Paper I is simply a
sequence of small events following a big one, similar to the small
oscillations in Figure 1.
Both the spectral evolution and the duration of the event are 
determined by one parameter only, namely the radius of the missing inner 
section of the accretion disk. For a large radius, the drop in
flux will be large and the time needed for re-fill will be long. 
Following Paper I, it is natural to associate the
length of the quiescent part of an event $t_q$ to the viscous time scale of the 
radiation-pressure dominated part of the accretion disk. This can
be expressed as $t_{\rm visc}=R^2/\nu$, where $\nu = \alpha c_S H$. 
From Frank, King \& Raine (1992) the scale-height $H$ and sound speed 
$c_S$ can be found, leading to 
\begin{equation}
t_{\rm visc} = 30 \alpha_2^{-1} M_1^{-1/2} R_7^{7/2}\dot M_{18}^{-2}\ {\rm s}
\end{equation}
where $\alpha_2 = \alpha/0.01$, $R_7$ is the radius in units of $10^7$ cm, 
$M_1$ the central object mass in solar masses, and $\dot M_{18}$ is the 
accretion rate in units of $10^{18}$ g/s. Notice that even the largest
radii derived here are well within the radiation-pressure dominated part
of the disk (see Equation 2 in Paper I).
The line in Figure 4 represents the best fit to the data with a relation
of the form $t_{\rm q} \propto R^{7/2}$. The fit is excellent, with
the exception of the point corresponding to the longest event. The
qualitative agreement with the theoretical expectation is striking, although
by substituting the appropriate values for the mass and accretion rate
we find that our best fit predicts rather small values of $\alpha_2$
(0.004 and 0.05 for the Schwarzschild and extreme Kerr cases
respectively).
This indicates a small viscosity in the disk, although we stress that 
$t_{visc}$ is only a time scale, so that additional corrections might
be necessary in order to allow a precise quantitative comparison.

An event can therefore be pictured in the following way.
At the start of a quiescent period, the disk has a central hole, whose
radius is R$_{max}$. The hole is either empty or filled with
gas whose radiation is too soft to be detected. Slowly the disk is re-filled
by a steady accretion rate $\dot M_0$ from outside. Each annulus of the disk
will move along the lower branch of its S-curve in the $\dot M-\Sigma$ plane 
trying to stabilize at $\dot M_0$ (see Paper I).
The surface gravity increases as the annulus moves towards the unstable
point at a speed determined by the local viscous time scale. 
During this period, no changes are observed in the radius of the hole,
since all the matter inside does not radiate in the PCA band. The observed
accretion rate is $\dot M_0$. At some point, 
one of the annuli will reach the unstable point and switch to the high-$\dot M$
state, where the accretion rate is larger than $\dot M_0$, causing a 
chain-reaction that will ``switch on'' the inner disk. The
observed accretion rate is now higher than the external value $\dot M_0$.
A smaller, hot radius is now observed. At the end of the outburst, the inner
disk runs out of fuel and switches off, either jumping back to the 
$\dot M < \dot M_0$ state or emptying completely. A new hole is formed and a 
new cycle starts.
Notice that in this scenario the more ``normal'' state for the source
is the one at high count rates, where the disk extends all the way to the
innermost stable orbit:
in this state the energy spectrum 
is similar to that of conventional black-hole candidates
(see \cite{tl95}). 

Not only the start and end of a major burst, but also all the small
amplitude oscillations within a burst show the same timing signature
of decaying faster than they rise. This is in agreement with what was already
noticed in Paper I: the rise time is determined by the speed at which
a heating wave moves through the central disk, while the faster decay time
is due to the rapid fall of matter into the black hole (or into a
relativistic jet).

Chen, Swank \& Taam (1997) found that when HR2 exceeds 0.1, 
the power density spectrum is similar to the one observed
in black-hole transient systems during the Very High State (see \cite{vdk95}).
In these occasions, a strong 1-6 Hz QPO peak was found, positively 
correlated with the count rate.
Notice that the limit HR2$>$0.1 is an indication that the source was
in a quiescent state. Our spectral results show that 
the fast timing features (both QPO and band-limited noise, see \cite{mrg96}) 
cannot originate from the innermost
regions of the optically thick accretion disk, 
since those are missing during the quiescent phases.
The fact that the QPO frequency increases with count rate is in qualitative
agreement with the model, since a higher count rate indicates a smaller
inner disk radius, and therefore shorter time scales.

The radii for the disappearing region of the disk found here are
considerably smaller than that reported in Paper I. This is
entirely due to the improved knowledge of the spectral response of the
PCA. During the burst phase the accretion rate through 
the optically thick accretion disk is
found to be higher than in the quiescent phase, contrary to what we reported
in Paper I.

Although we can model the spectral evolution and the time scale of
the events, most of the variability is yet to be explained. The main
question to be answered is: what determines the length of the
following outburst? Or, in more physical terms, what determines how large
the next missing section of the disk will be? In some observations the
events are very regular, in some others they are extremely random, and in
some others no events are observed at all. In the observation reported
here a striking one to one relation between quiescent and burst time is
observed, a relation which 
applies to the ``outburst'' and ``quiescent'' states of the observation
presented in Paper I but
is obviously not satisfied in other observations
(see e.g. \cite{tcs97}) 
nor during the ``flare'' state of Paper I.
Moreover, as already mentioned, one observation among the ones we analyzed
does not fit this pattern and requires a different interpretation 
(1996 June 16th).  Nevertheless, the model
sketched here provides a satisfactory interpretation of the cause of
the changes in the X-ray emission. It fits not only the energy spectral
distribution, but also the complex variability shown by GRS~1915+105.

\acknowledgements

MM is a fellow of the Consejo Nacional de Investigaciones
Cient\'{\i}ficas y T\'ecnicas de la Rep\'ublica Argentina.
This work was supported in part by the Netherlands Organization for
Scientific Research (NWO) under grant PGS 78-277 and the Netherlands
Foundation for Research in Astronomy (ASTRON) under grant 781-76-017.
ARK gratefully acknowledges a PPARC Senior Fellowship. 
JVP acknowledges
support from NASA under contract NAG 5-3003. We thank Keith Jahoda for
helping with the calibration of the PCA data, and the RXTE team for making
these data publicly available. We thank D. Battacharya for his careful reading
of the manuscript.

{}

\clearpage

\figcaption[]{Upper panel: 2.0--40.0 keV PCA light curve.
             Time zero corresponds to 1997 June 18th 14:36 UT.
             Middle and lower panels: corresponding inner radius
             and temperature (see text).
            }

\figcaption[]{Correlation between burst and quiescent time (as
	     defined in the text). The line is a y=x relation.
            }
\figcaption[]{Three color-color diagrams for events grouped by total length.}

\figcaption[]{Correlation between total length of the event and maximum
            inner radius of the disk. The line is the best fit with a
	    power law with fixed index $\gamma = 3.5$ and normalization
            1.44$\times 10^{-5}$. The last point
            has been excluded from the fit.}

\clearpage

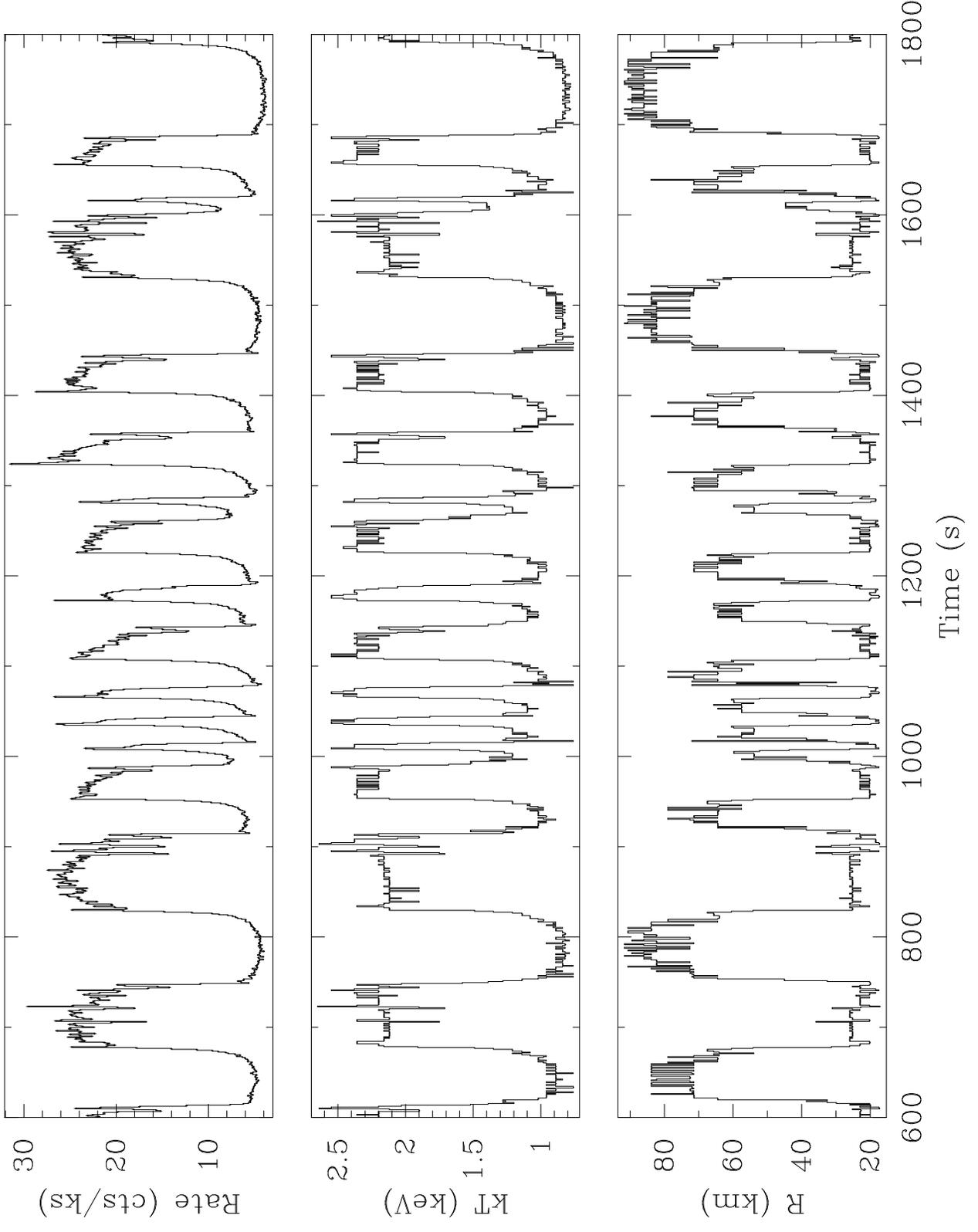
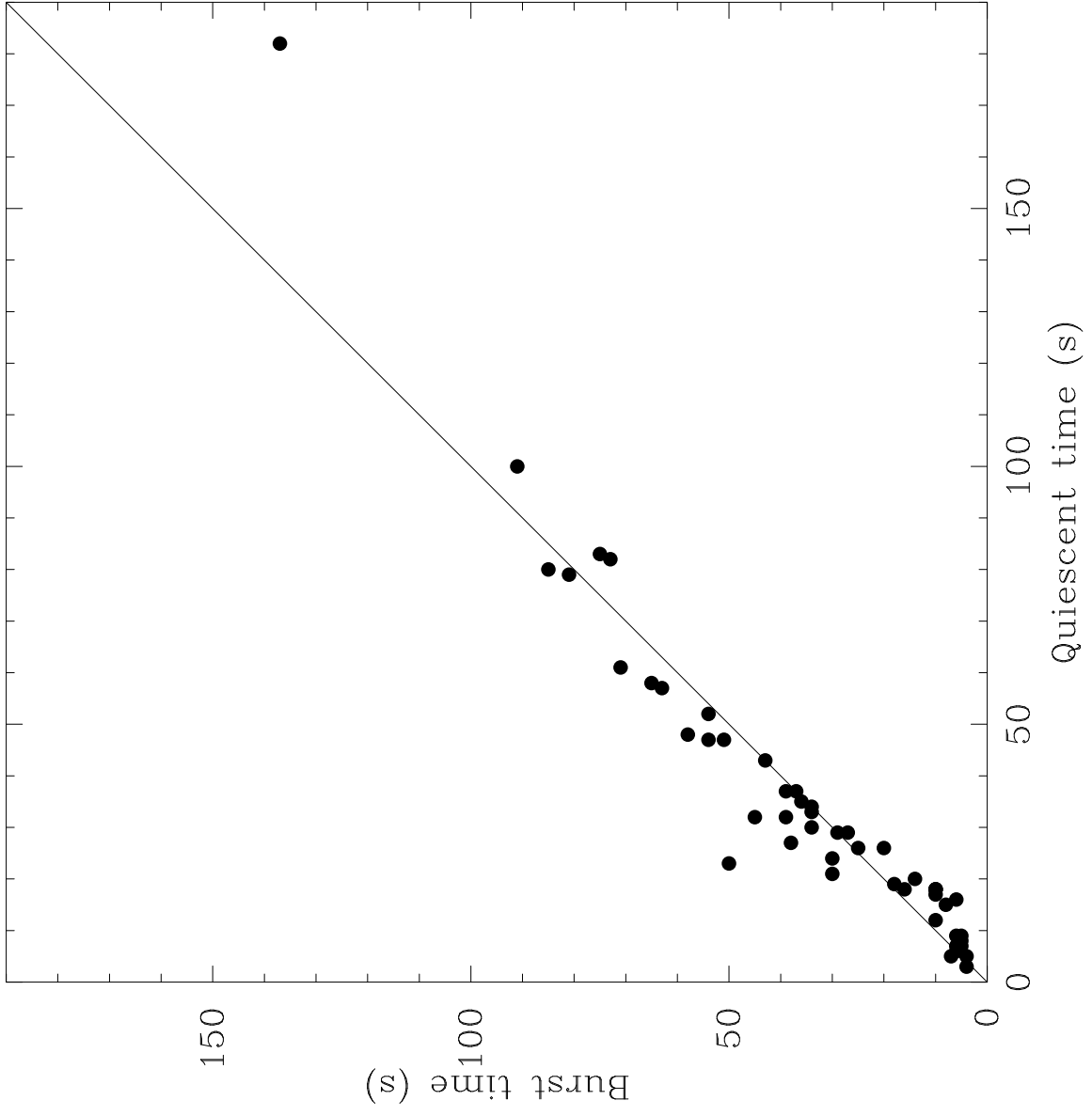
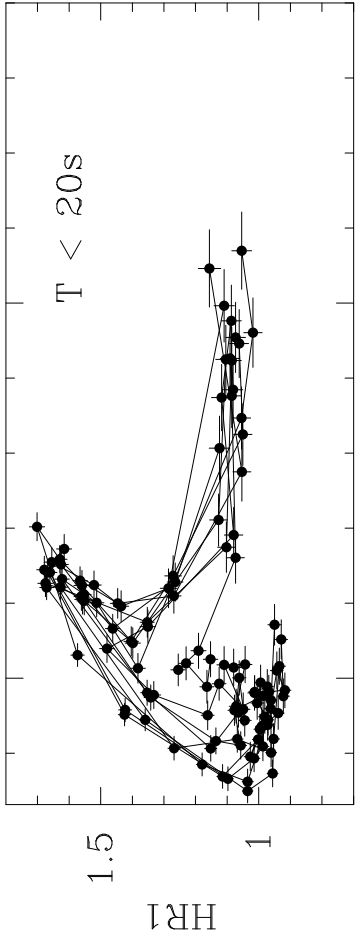
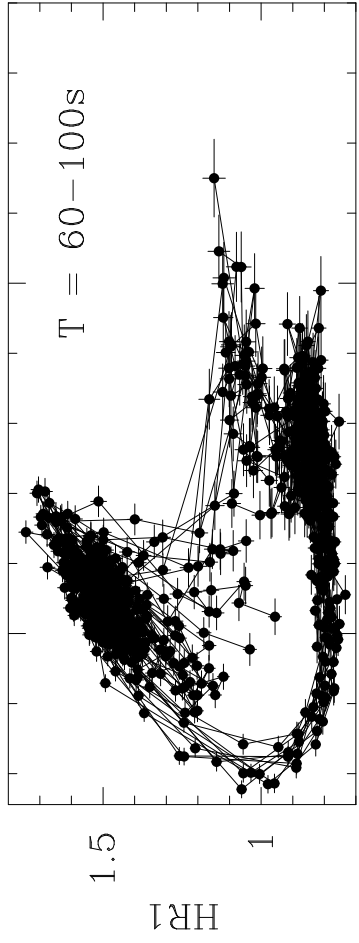
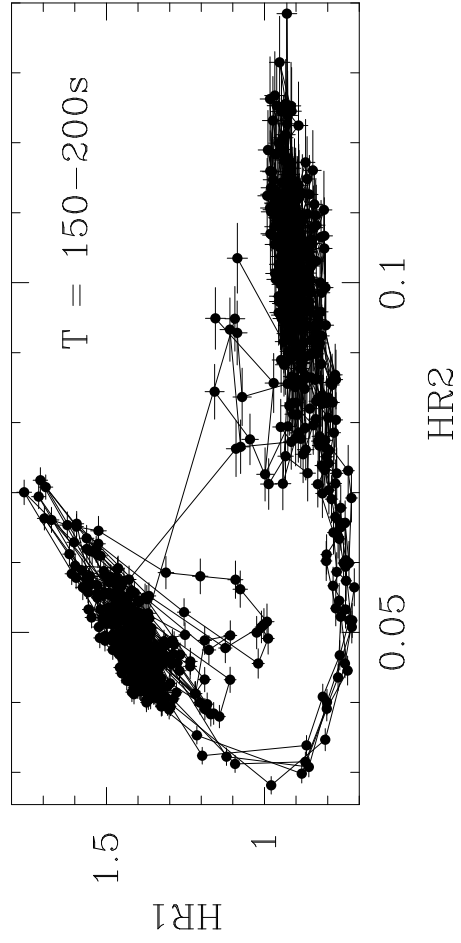
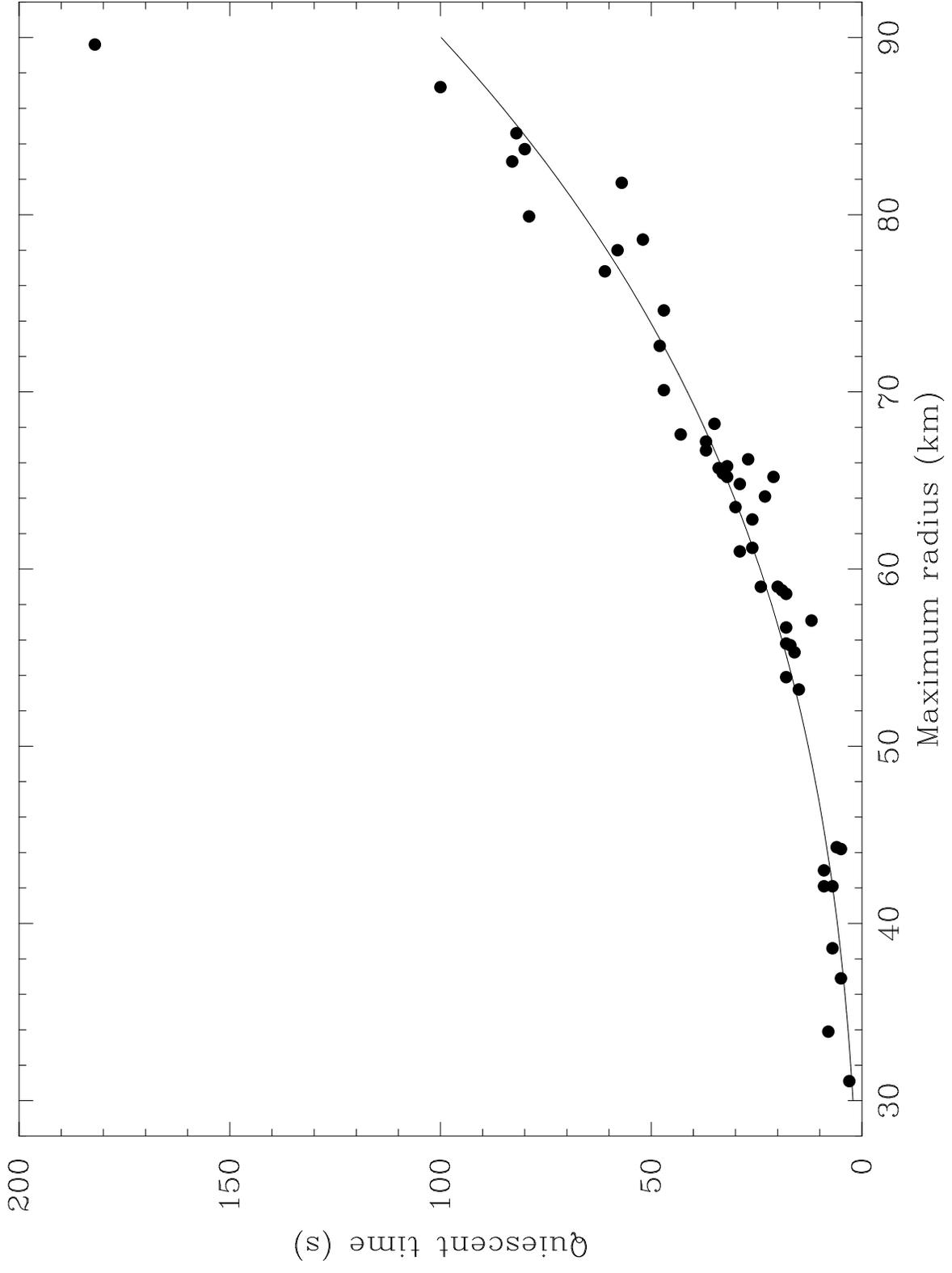
\end{document}